\documentclass[12pt,doublespacing]{article}

\usepackage{fullpage}
\usepackage{setspace}
\usepackage{authblk}

\usepackage{cite}
\usepackage{hyperref} 

\usepackage{graphicx}

\usepackage{amsmath}
\usepackage{amssymb}
\allowdisplaybreaks

\usepackage{float}
\usepackage{subfig}
\usepackage{makecell}
\usepackage{rotating}

\usepackage{enumitem}

\newcounter{manuscriptalgorithm}

\begin{document}

\title{A Sparsity-Promoting Electromagnetic Inversion Method Regularized by $\ell_1 / \ell_2$-Norm of the Model Gradient}

\author[]{Lingqi Gao}
\author[]{Hakan Bagci\vspace{0.5cm}}

\affil[]{Electrical and Computer Engineering (ECE) Program Computer
\authorcr Electrical, and Mathematical Science and Engineering (CEMSE) Division
\authorcr King Abdullah University of Science and Technology (KAUST)
\authorcr Thuwal, 23955-6900, Saudi Arabia
\authorcr e-mails:\{lingqi.gao@kaust.edu.sa hakan.bagci@kaust.edu.sa\}}

\date{}
\maketitle
\newpage

\begin{abstract}
A nonlinear EM inversion method regularized by $\ell_1 / \ell_2$-norm of the model gradient is proposed. The $\ell_1 / \ell_2$-norm, defined as the ratio of $\ell_1$-norm to $\ell_2$-norm, exhibits a scale-invariant property that enables it to more accurately characterize sparsity in the solution compared with the conventional $\ell_1$-norm. To address the nonconvexity and nonsmoothness introduced by the quotient structure of $\ell_1 / \ell_2$-norm, the inversion problem is reformulated and solved using the alternating direction method of multipliers, resulting in an efficient optimization algorithm comprising five sub-steps. The Gauss--Newton method is incorporated into the first sub-step to linearize the nonlinear inversion. The proposed method is validated by a series of numerical examples using both synthetic and experimental datasets, and is compared against standard regularization methods such as Tikhonov and total variation. The results demonstrate that the proposed method yields superior reconstructions of permittivity profiles, particularly in preserving sharp edges and suppressing artifacts, without incurring additional computational cost. Furthermore, the method exhibits strong robustness to measurement noise and reduced sensitivity to the choice of the regularization weight.
\end{abstract}

\section{Introduction}\label{introduction}

Electromagnetic (EM) inversion aims to reconstruct the unknown electrical properties such as permittivity and/or conductivity within a domain of interest using scattered field measurements collected externally~\cite{Microwave Imaging}. Because of its low cost and non-invasive nature, EM inversion has been successfully applied to various fields such as subsurface imaging~\cite{Ground Penetrating Radar}, through-wall imaging~\cite{Through-wall imaging (TWI) by radar: 2-D tomographic results and analyses}, and biomedical imaging~\cite{Stroke diagnosis using microwave techniques: Review of systems and algorithms}.

Over the past few decades, various EM inversion algorithms have been successfully developed. Simpler linearized approaches, such as those based on the Born approximation, are computationally efficient but produce accurate reconstructions only when the scatterers are weak. To partially mitigate this limitation and improve the handling of nonlinearity, several extensions have been proposed, including the Born iterative method (BIM)~\cite{Microwave imaging for brain stroke detection using Born iterative method, Variational Born iteration method and its applications to hybrid inversion}, higher-order Born approximation~\cite{An inexact-Newton method for short-range microwave imaging within the second-order Born approximation}, and the extended Born approximation~\cite{Two nonlinear inverse methods for electromagnetic induction measurements}. More advanced methods that account for the full nonlinearity have also been proposed. These methods do not rely on any approximation for linearization and, in principle, can reconstruct stronger scatterers, albeit at a significantly higher computational cost. They primarily include the contrast source inversion (CSI)~\cite{A contrast source inversion method}, the distorted Born iterative method (DBIM)~\cite{Reconstruction of two-dimensional permittivity distribution using the distorted Born iterative method}, and optimization-based inversion algorithms~\cite{Conjugate gradient method applied to inverse scattering problem, Microwave imaging-complex permittivity reconstruction with a Levenberg-Marquardt method}.

In addition to nonlinearity, EM inversion is inherently ill-posed, making the reconstruction process highly sensitive to noise and modeling errors. To address ill-posedness, regularization techniques are employed to incorporate prior information about the solution and stabilize the inversion process~\cite{Parameter Estimation and Inverse Problems}. Among them, Tikhonov regularization~\cite{Tikhonov regularization and total least squares} is one of the most widely used methods in EM inversion~\cite{Microwave imaging for brain stroke detection using Born iterative method, A two-stage microwave image reconstruction procedure for improved internal feature extraction, Overview and classification of some regularization techniques for the Gauss-Newton inversion method applied to inverse scattering problems}. It promotes smoothness in the reconstructed solution by minimizing the square of the $\ell_2$-norm of the model. An alternative approach involves subspace projection, which can be implemented by early stopping of iterative solvers such as the Landweber or conjugate gradient methods~\cite{Nonlinear microwave imaging for breast-cancer screening using Gauss–Newton's method and the CGLS inversion algorithm, A convergence analysis of the Landweber iteration for nonlinear ill-posed problems}. While Tikhonov and subspace projection methods are effective in suppressing instability in the inversion, they often result in an over-smoothed reconstruction, especially in the presence of abrupt spatial variations.

However, many EM inversion applications demand high spatial resolution in the reconstructed profiles, making over-smoothing undesirable. This motivates the use of sparsity-promoting regularization strategies, which enforce sparsity in the solution itself or in a transformed domain, such as the gradient domain [e.g., total variation (TV)] or the wavelet domain~\cite{Sparsity regularized nonlinear inversion for microwave imaging, Shrinkage-thresholding enhanced Born iterative method for solving 2D inverse electromagnetic scattering problem, A preconditioned inexact Newton method for nonlinear sparse electromagnetic imaging, Non-linear inverse scattering via sparsity regularized contrast source inversion, A sparsity-regularized Born iterative method for reconstruction of two-dimensional piecewise continuous inhomogeneous domains, Limited-angle CT reconstruction via the L1/L2 minimization, A sparsity regularization approach to the electromagnetic inverse scattering problem, Extended contrast source inversion}. The most direct way to impose such sparsity constraints is to include $\ell_0$-norm as a regularization term, as it explicitly counts the number of nonzero elements in the solution~\cite{Shrinkage-thresholding enhanced Born iterative method for solving 2D inverse electromagnetic scattering problem}. However, due to the nonconvex and combinatorial nature of $\ell_0$-norm minimization, it is often replaced by $\ell_1$-norm, which serves as a convex surrogate.

The $\ell_1$-norm is convex and only non-differentiable at zero, making the resulting optimization problem significantly easier to solve. The $\ell_1$-norm regularization has been successfully incorporated into various EM inversion algorithms, including BIM~\cite{Sparsity regularized nonlinear inversion for microwave imaging} and CSI~\cite{A preconditioned inexact Newton method for nonlinear sparse electromagnetic imaging}. Additionally, applying $\ell_1$-norm to the gradient of the solution imposes a piecewise-constant prior, widely known as TV regularization. TV has been adopted in several EM inversion algorithms, such as CSI~\cite{Non-linear inverse scattering via sparsity regularized contrast source inversion} and BIM~\cite{A sparsity-regularized Born iterative method for reconstruction of two-dimensional piecewise continuous inhomogeneous domains}. The use of $\ell_1$-norm-based regularization often leads to sharper and more accurate reconstructions compared with $\ell_2$-norm-based regularization~\cite{Sparsity regularized nonlinear inversion for microwave imaging, Shrinkage-thresholding enhanced Born iterative method for solving 2D inverse electromagnetic scattering problem, A preconditioned inexact Newton method for nonlinear sparse electromagnetic imaging, Non-linear inverse scattering via sparsity regularized contrast source inversion, A sparsity-regularized Born iterative method for reconstruction of two-dimensional piecewise continuous inhomogeneous domains, Limited-angle CT reconstruction via the L1/L2 minimization, A sparsity regularization approach to the electromagnetic inverse scattering problem, Extended contrast source inversion}. However, it is also known to introduce artifacts such as the staircasing effect and the underestimation bias~\cite{Limited-angle CT reconstruction via the L1/L2 minimization}. In addition, other regularization methods have been proposed to promote sparsity, such as the elastic net, which combines $\ell_1$- and $\ell_2$-norm penalties~\cite{A sparsity regularization approach to the electromagnetic inverse scattering problem}, and multiplicative regularization, which introduces TV as a multiplicative term in the minimization problem~\cite{Extended contrast source inversion}.

The quality of sparsity-promoting inversion can further be improved by employing regularization functions that more accurately approximate $\ell_0$-norm while yielding optimization problems that can be solved efficiently. Beyond the methods discussed above, an increasingly popular approach involves using the ratio of $\ell_1$-norm to $\ell_2$-norm, known as $\ell_1 / \ell_2$-norm, which is particularly appealing due to its scale-invariant property in measuring sparsity~\cite{A scale-invariant approach for sparse signal recovery}. The idea of $\ell_1 / \ell_2$-norm was originally introduced in~\cite{Non-negative matrix factorization with sparseness constraints} for non-negative matrix factorization and later investigated in~\cite{Comparing measures of sparsity} as a measure of  sparsity. Since then, $\ell_1 / \ell_2$-norm has been widely adopted as a sparsity-promoting regularization in various applications, including compressed sensing~\cite{Estimating unknown sparsity in compressed sensing, Minimization of L1 over L2 for sparse signal recovery with convergence guarantee}, blind deconvolution~\cite{Blind deconvolution using a normalized sparsity measure}, and image reconstruction in computed tomography (CT) and magnetic resonance imaging (MRI)~\cite{Limited-angle CT reconstruction via the L1/L2 minimization, Minimizing L1 over L2 norms on the gradient}. For inverse problems, it has been applied directly to the model as well as to its gradient domain~\cite{Limited-angle CT reconstruction via the L1/L2 minimization, Minimizing L1 over L2 norms on the gradient} and wavelet domain representations~\cite{Image deconvolution using a characterization of sharp images in wavelet domain}.

Despite its appeal, $\ell_1 / \ell_2$-norm is difficult to use as a regularizer, because the resulting optimization problem is nonconvex and nonsmooth. For instance, in~\cite{Blind deconvolution using a normalized sparsity measure}, the denominator is fixed during optimization based on the previous iterate, reducing the problem to a standard $\ell_1$-norm regularization. The alternating direction method of multipliers (ADMM)~\cite{Distributed optimization and statistical learning via the alternating direction method of multipliers} offers an alternative by decomposing the $\ell_1 / \ell_2$-norm-regularized problem into a sequence of more manageable sub-problems~\cite{Limited-angle CT reconstruction via the L1/L2 minimization, A scale-invariant approach for sparse signal recovery, Minimizing L1 over L2 norms on the gradient}.

More recently, other quotient-based norms have also been proposed. For example, a smoothed $\ell_1 / \ell_2$-norm was introduced in~\cite{Euclid in a Taxicab: sparse blind deconvolution with smoothed l1/l2 regularization} to mitigate the nonconvexity and nonsmoothness of the original $\ell_1 / \ell_2$-norm. This idea was later generalized to a broader class of $\ell_p / \ell_q$-norms in~\cite{SPOQ lp/lq regularization for sparse signal recovery applied to mass spectrometry}. In~\cite{A proximal algorithm with backtracked extrapolation for a class of structured fractional programming}, a $\ell_{1} / s_{K}$-norm was proposed as a sparsity-promoting function, where $s_{K}$ represents the sum of the $K$ largest absolute values in a vector. These variants of $\ell_1 / \ell_2$-norm regularization introduce additional parameters, which are often difficult to tune effectively in inversion problems.

In this paper, $\ell_1/\ell_2$-norm of the model gradient is adopted as a sparsity-promoting regularization for nonlinear EM inversion. The resulting optimization problem is solved using ADMM, which is designed with five sub-steps to handle the quotient structure of $\ell_1/\ell_2$-norm. The Gauss--Newton method is incorporated into the first sub-step of the ADMM framework to linearize the nonlinear inversion problem, and a frequency-hopping scheme, as described in~\cite{A frequency-hopping approach for microwave imaging of large inhomogeneous bodies}, is employed to enable accurate, robust, and computationally efficient multi-frequency inversion. The results demonstrate that the proposed method overcomes the shortcomings of existing $\ell_1$-norm-based regularization methods, providing cleaner reconstructions with sharper edges. This improvement stems from the scale-invariant property of $\ell_1 / \ell_2$-norm. A preliminary version of this work was reported in~\cite{A ratio-norm regularization for sparsity-promoting electromagnetic inversion}.

The rest of this paper is organized as follows. In Section~\ref{formulation}, the forward problem is first derived, followed by the mathematical formulation of nonlinear EM inversion regularized by $\ell_1/\ell_2$-norm, where the sparsity-promoting behavior of $\ell_1 / \ell_2$-norm is illustrated through a 2D visualization. In Section~\ref{numerical_examples}, numerical examples with synthetic and experimental data are presented. Section~\ref{conclusions} concludes this work with a brief summary.

\section{Formulation}\label{formulation}
\subsection{Forward Problem}\label{forward_problem}
\begin{figure}[!ht]
\centering
\includegraphics[width=0.6\columnwidth]{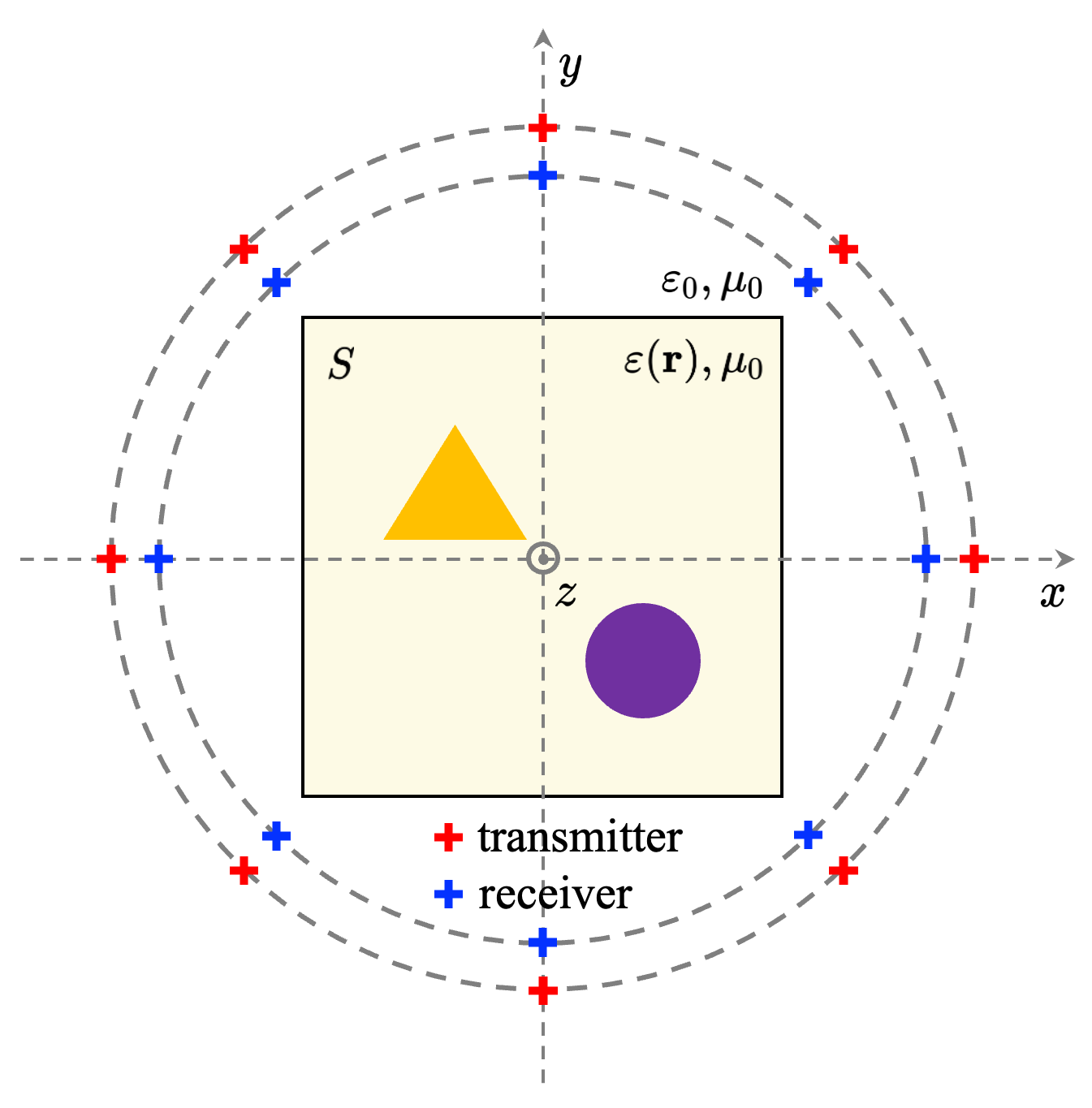}
\caption{Schematic of the 2D inverse scattering configuration, showing the investigation domain $S$ surrounded by transmitters and receivers.}\label{system_configuration}
\end{figure}

Let $S$ represent an investigation domain embedded in an unbounded background with permittivity $\varepsilon_0$ and permeability $\mu_0$. The problem is assumed to be $z$-invariant, i.e., two-dimensional (2D), with the position vector expressed in the $xy$-plane as $\mathbf{r} = x\hat{\mathbf{x}}+y\hat{\mathbf{y}}$. Within $S$, the permittivity is $\varepsilon(\mathbf{r})$, while the permeability remains $\mu_0$. The domain $S$ is surrounded by $N_{\mathrm{t}}$ transmitters and $N_{\mathrm{r}}$ receivers (see Fig.~\ref{system_configuration}). $S$ is illuminated by one transmitter at a time, and the scattered fields are recorded by the receivers for each illumination. The dielectric contrast is defined as
\begin{equation}\label{contrast_definition}
\tau(\mathbf{r})= \begin{cases}\varepsilon(\mathbf{r}) / \varepsilon_0-1, & \mathbf{r} \in S \\ 0, & \mathrm{else}\end{cases}.
\end{equation}
For illumination by transmitter $i$, $i=1,2,\ldots,N_{\mathrm{t}}$,  the 2D transverse magnetic (TM) scattering problem can be described by the following equations
\begin{equation}
\label{scattered_field_equation}
E_i^{\mathrm{sca}}(\mathbf{r})=-k_0^2 \int_S \tau(\mathbf{r}^{\prime}) E_i^{\mathrm{tot}}(\mathbf{r}^{\prime}) g(\mathbf{r}, \mathbf{r}^{\prime}) d s^{\prime}
\end{equation}
\begin{equation}
\label{total_field_equation}
\begin{aligned}
E_{i}^{\mathrm{inc}}(\mathbf{r}) & =E_{i}^{\mathrm{tot}}(\mathbf{r}) - E_{i}^{\mathrm{sca}}(\mathbf{r})\\
E_{i}^{\mathrm{inc}}(\mathbf{r}) &= E_{i}^{\mathrm{tot}}(\mathbf{r})+k_0^2 \int_S \tau(\mathbf{r}^{\prime}) E_i^{\mathrm{tot}}(\mathbf{r}^{\prime}) g(\mathbf{r}, \mathbf{r}^{\prime}) d s^{\prime}.
\end{aligned}
\end{equation}
Here, $E_i^{\mathrm{inc}}(\mathbf{r})$ is the $z$-component of the incident electric field generated by transmitter $i$, and $E_i^{\mathrm{sca}}(\mathbf{r})$ and $E_i^{\mathrm{tot}}(\mathbf{r})$  are the $z$-components of the scattered and total electric fields, respectively. The background medium wavenumber is given by $k_0=\omega\sqrt{\mu_0 \varepsilon_0} = 2\pi/\lambda_0$, where $\omega$ is the angular frequency and $\lambda_0$ is the free-space wavelength. The 2D scalar Green function is expressed as $g(\mathbf{r}, \mathbf{r}^{\prime})=(\mathrm{j}/{4}) H_0^{(2)}(k_0|\mathbf{r}-\mathbf{r}^{\prime}|)$, where $H_0^{(2)}(\cdot)$ denotes the zeroth-order Hankel function of the second kind.

To discretize~\eqref{scattered_field_equation} and~\eqref{total_field_equation}, $S$ is divided into $N$ square cells, and $\tau(\mathbf{r})$ and $E_i^{\mathrm{tot}}(\mathbf{r})$ are expanded as
\begin{equation}\label{field_expansions}
\begin{aligned}
\tau(\mathbf{r})&=\sum_{n=1}^N\{\bar{\tau}\}_n p_n(\mathbf{r})\\
E_i^{\mathrm{tot}}(\mathbf{r})&=\sum_{n=1}^N\{\bar{E}_i^{\mathrm{tot}}\}_n p_n(\mathbf{r}).
\end{aligned}
\end{equation}
Here, $\{\bar{\tau}\}_n = \tau(\mathbf{r}_n)$ and $\{\bar{E}_i^{\mathrm{tot}}\}_n = E_i^{\mathrm{tot}}(\mathbf{r}_n)$, $n = 1,2,\ldots,N$, where $\{\cdot\}_n$ denotes the $n$-th entry of a vector and $\mathbf{r}_n$ denotes the center of cell $n$.  The pulse basis function $p_n(\mathbf{r})$ is given by
\begin{equation}\label{pulse_basis}
p_n(\mathbf{r})= \begin{cases}1, & \mathbf{r} \in S_n \\ 0, & \mathrm{else}\end{cases}
\end{equation}
where $S_n$ is the support of cell $n$. Inserting~\eqref{field_expansions} into~\eqref{total_field_equation} and sampling the resulting equation at $\mathbf{r}_m$, $m=1,2,\ldots,N$ yields the following matrix equation
\begin{equation}\label{discrete_forward_system}
[\bar{\bar{I}}+\bar{\bar{G}} \mathrm{diag}\{\bar{\tau}\}] \bar{E}^{\mathrm{tot}}_{i}=\bar{E}^{\mathrm{inc}}_{i}.
\end{equation}
Here, $\bar{\bar{I}}$ represents the identity matrix, $\mathrm{diag} \{ \cdot \}$ is an operator that converts a vector into a diagonal matrix, and $\bar{\bar{G}}$ is the domain Green function matrix with entries given by
\begin{equation}
\label{domain_green_matrix}
\{\bar{\bar{G}}\}_{mn}=k_0^2\int_{S_n} g(\mathbf{r}_m, \mathbf{r}^{\prime})\,d s^{\prime}, \; m,n = 1,2,\ldots,N
\end{equation}
where $\{\cdot\}_{mn}$ denotes the entry in the $m$-th row and $n$-th column of a matrix. Let $\mathbf{r}_j^{\mathrm{r}}$ and $\{\bar{E}^{\mathrm{sca}}_{i}\}_j = E^{\mathrm{sca}}_{i}(\mathbf{r}_j^{\mathrm{r}})$, $j=1,2,\ldots,N_{\mathrm{r}}$, represent the locations of the receivers and the samples of the scattered field at these locations.  Inserting~\eqref{field_expansions} into~\eqref{scattered_field_equation} and sampling the resulting equation at $\mathbf{r}_j^{\mathrm{r}}$  yields the expression for $\bar{E}^{\mathrm{sca}}_{i}$ as
\begin{equation}\label{predicted_scattered_field}
\bar{E}^{\mathrm{sca}}_{i}=\bar{\bar{H}} \mathrm{diag}\{\bar{E}^{\mathrm{tot}}_{i}\} \bar{\tau}
\end{equation}
where $\bar{\bar{H}}$ is the receiver Green function matrix with entries given by
\begin{equation}\label{receiver_green_matrix}
\{\bar{\bar{H}}\}_{jn}=-k_0^2\int_{S_n} g(\mathbf{r}_j^{\mathrm{r}}, \mathbf{r}^{\prime})\,d s^{\prime}, \; j = 1,2,\ldots,N_{\mathrm{r}},\;n = 1,2,\ldots,N.
\end{equation}
For a given $\bar{\tau}$, first~\eqref{discrete_forward_system} is solved for $\bar{E}^{\mathrm{tot}}_{i}$, then $\bar{E}^{\mathrm{tot}}_{i}$ is used in~\eqref{predicted_scattered_field} to compute $\bar{E}^{\mathrm{sca}}_{i}$. Together,~\eqref{discrete_forward_system} and~\eqref{predicted_scattered_field} constitute the forward problem, i.e., the mapping from a given $\bar{\tau}$ to the scattered field $\bar{E}^{\mathrm{sca}}_{i}$. In this work,~\eqref{discrete_forward_system} is solved using a direct solver.

\subsection{Inverse Problem}\label{inverse_problem}
The contrast $\bar{\tau}$ is reconstructed by minimizing the following objective function
\begin{equation}\label{inversion_objective}
\Phi(\bar{\tau})=F(\bar{\tau})+\lambda R(\bar{\tau})=\frac{1}{2}\left\|\bar{E}^{\mathrm{mea}}-\bar{E}^{\mathrm{prd}}\right\|_2^2+\lambda R(\bar{\tau})
\end{equation}
where $\bar{E}^{\mathrm{prd}} = [\bar{E}^{\mathrm{prd},\top}_{1}, \bar{E}^{\mathrm{prd},\top}_{2}, \ldots , \bar{E}^{\mathrm{prd},\top}_{N_{\mathrm{t}}}]^{\top}$ collects the scattered fields predicted at the receiver locations for all transmitters, with each vector $\bar{E}^{\mathrm{prd}}_{i}$ computed
from~\eqref{predicted_scattered_field} for illumination by transmitter $i$ using a guess $\bar{\tau}$, while $\bar{E}^{\mathrm{mea}} = [\bar{E}^{\mathrm{mea},\top}_{1}, \bar{E}^{\mathrm{mea},\top}_{2}, \ldots , \bar{E}^{\mathrm{mea},\top}_{N_{\mathrm{t}}}]^{\top}$ collects the corresponding measured scattered fields. Here, ``$\top$'' denotes the transpose operation. $F(\bar{\tau})$ is the data misfit term measuring the squared $\ell_2$-norm distance between the measured data and the predicted data, and $R(\bar{\tau})$ is the regularization term, weighted by $\lambda$, that encodes prior information about the solution. To promote the sparsity of the model gradient, $\ell_0$-norm and $\ell_1$-norm are often used, resulting in the following objective functions
\begin{equation}\label{l0_objective}
\Phi(\bar{\tau})=\frac{1}{2}\left\|\bar{E}^{\mathrm{mea}}-\bar{E}^{\mathrm{prd}}\right\|_2^2+\lambda\|\bar{\bar{D}} \bar{\tau}\|_0
\end{equation}
\begin{equation}\label{tv_objective}
\Phi(\bar{\tau})=\frac{1}{2}\left\|\bar{E}^{\mathrm{mea}}-\bar{E}^{\mathrm{prd}}\right\|_2^2+\lambda\|\bar{\bar{D}} \bar{\tau}\|_1
\end{equation}
where $\bar{\bar{D}}$ is the discretized gradient operator, so that $\bar{\bar{D}}\bar{\tau}$ approximates the spatial gradient of the contrast and $\|\bar{\bar{D}}\bar{\tau}\|_1$ is the TV of the contrast. Although $\ell_0$-norm directly describes the number of nonzero elements in the model gradient,~\eqref{l0_objective} is hard to optimize because the combinatorial nature of $\ell_0$-norm minimization makes it NP-hard~\cite{Sparse approximate solutions to linear systems}. Instead, the well-known TV regularization~\eqref{tv_objective} offers a convex approximation to~\eqref{l0_objective} by replacing $\ell_0$-norm with $\ell_1$-norm. However, $\ell_1$-norm is absolutely homogeneous: for any scalar $c$, $\|c\bar{x}\|_1 = |c|\,\|\bar{x}\|_1$, so its value scales linearly with the overall magnitude of the vector. In a regularized objective, the penalty can be reduced by shrinking the reconstructed values, which biases the solution toward underestimated magnitudes. Normalizing $\ell_1$-norm by $\ell_2$-norm removes this magnitude dependence: since both norms are homogeneous of degree one, the ratio $\|c\bar{x}\|_1/\|c\bar{x}\|_2 = \|\bar{x}\|_1/\|\bar{x}\|_2$ is scale-invariant. The resulting $\ell_1 / \ell_2$ ratio (commonly referred to as $\ell_1 / \ell_2$-norm) thus measures sparsity independently of magnitude, avoiding the underestimation bias of $\ell_1$-norm.

\begin{figure}[!ht]
\centering
\includegraphics[width=1\columnwidth]{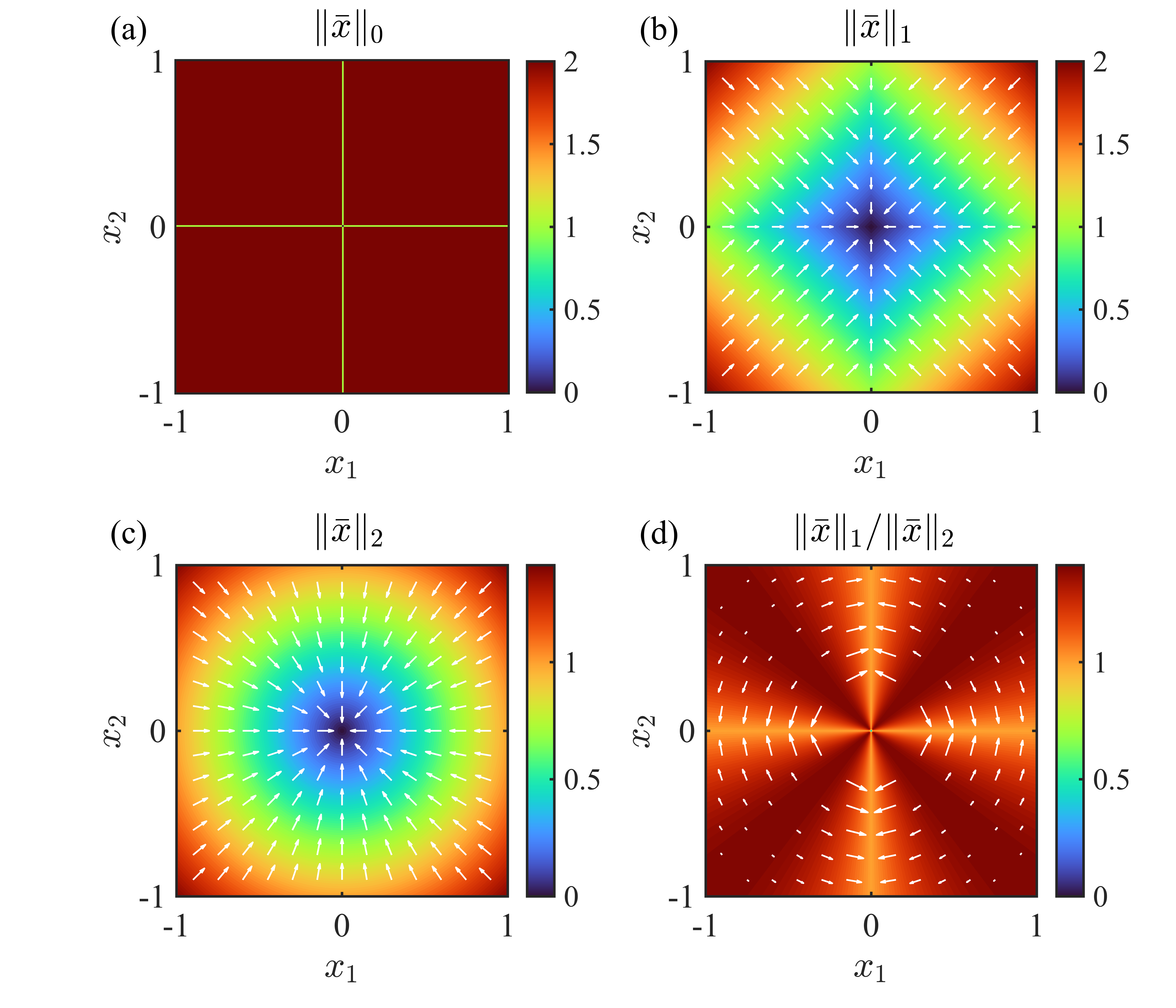}
\caption{2D visualization of (a) $\ell_0$-norm, (b) $\ell_1$-norm, (c) $\ell_2$-norm, (d) $\ell_1 / \ell_2$-norm. The white arrows in (b)--(d) indicate the direction of the negative gradient.}\label{norm_visualization}
\end{figure}

Fig.~\ref{norm_visualization} compares the four norms for a 2D vector $\bar{x} = [x_1, x_2]$, where each colormap shows the norm value and the white arrows show the negative-gradient direction, i.e., the direction along which optimization drives the solution. The $\ell_0$-norm is the ideal sparsity measure: it equals $2$ where both components are nonzero, drops to $1$ on the coordinate axes where one component vanishes, and equals $0$ at the origin [Fig.~\ref{norm_visualization}(a)]. However, it is piecewise constant, so its gradient is zero wherever it is defined, and it provides no descent direction to optimize [hence the absence of arrows in Fig.~\ref{norm_visualization}(a)]. The $\ell_2$-norm is smooth, but its level sets are circles and its negative gradient is radial, pointing directly at the origin [Fig.~\ref{norm_visualization}(c)]. Optimization therefore moves $\bar{x}$ along its own ray, shrinking both components by the same factor while leaving their ratio unchanged. The magnitude decreases, but no component is ever preferentially driven to $0$, so $\ell_2$ does not promote sparsity. The $\ell_1$-norm, the standard convex surrogate for $\ell_0$-norm, does promote sparsity: its negative gradient points toward the axes, decreasing both components at equal rates [Fig.~\ref{norm_visualization}(b)], so the smaller component reaches $0$ first. But the penalty keeps decreasing with magnitude, so the flow does not stop at the axis. It continues to the origin, driving the surviving component down and eventually to $0$. The downward pull on the nonzero component is the origin of the underestimation bias of $\ell_1$-norm. The $\ell_1/\ell_2$-norm removes this bias while retaining the sparsifying behavior [Fig.~\ref{norm_visualization}(d)]. Over all nonzero vectors its value lies in $[1, \sqrt{2}]$: it attains its minimum of $1$ on the coordinate axes, and its maximum of $\sqrt{2}$ along the diagonals where $|x_1| = |x_2|$ (at the origin it is assigned the value $0$ by the convention $\|\bar{0}\|_1/\|\bar{0}\|_2 = 0$). This landscape mirrors that of $\ell_0$-norm, with the axes forming the valleys. Because the ratio is scale-invariant, i.e., homogeneous of degree zero, its gradient is orthogonal to $\bar{x}$. The negative gradient is therefore tangential [Fig.~\ref{norm_visualization}(d)]: it rotates $\bar{x}$ toward the nearest axis at essentially constant magnitude, driving the smaller component to $0$ without shrinking the vector. In this way, $\ell_1/\ell_2$-norm reproduces the sparsity-promoting property of $\ell_0$-norm while supplying a usable descent direction, and it sparsifies the solution without the magnitude collapse that biases $\ell_1$-norm. 

With $\ell_1 / \ell_2$-norm regularization on the model gradient, the new objective  is
\begin{equation}\label{ratio_norm_objective}
\Phi(\bar{\tau})=\frac{1}{2}\left\|\bar{E}^{\mathrm{mea}}-\bar{E}^{\mathrm{prd}}\right\|_2^2+\lambda \frac{\|\bar{\bar{D}} \bar{\tau}\|_1}{\|\bar{\bar{D}} \bar{\tau}\|_2}.
\end{equation}
Since the $\ell_1 / \ell_2$-norm regularization is nonconvex and nonsmooth, \eqref{ratio_norm_objective} is not straightforward to optimize. Instead, in this paper, ADMM~\cite{Distributed optimization and statistical learning via the alternating direction method of multipliers} is used to decouple the numerator and the denominator of the regularization term so that they can be updated separately. To this end, two auxiliary variables subject to $\bar{n}=\bar{\bar{D}} \bar{\tau}$ and $\bar{p}=\bar{\bar{D}} \bar{\tau}$ are introduced so that the optimization of~\eqref{ratio_norm_objective} becomes
\begin{equation}\label{constrained_ratio_problem}
\bar{\tau} = \arg \min _{\bar{\tau}} \frac{1}{2}\left\|\bar{E}^{\mathrm{mea}}-\bar{E}^{\mathrm{prd}}\right\|_2^2+\lambda \frac{\|\bar{n}\|_1}{\|\bar{p}\|_2} \quad \mathrm{s.t.}\; \bar{n}=\bar{\bar{D}} \bar{\tau}, \; \bar{p}=\bar{\bar{D}} \bar{\tau}.
\end{equation}
Although both auxiliary variables equal $\bar{\bar{D}}\bar{\tau}$, keeping them separate allows the $\ell_1$ numerator and the $\ell_2$ denominator to be updated independently. The corresponding augmented Lagrangian of~\eqref{constrained_ratio_problem} reads
\begin{equation}\label{augmented_lagrangian}
\begin{aligned}
L(\bar{\tau}, \bar{n}, \bar{p}, \bar{u}, \bar{q}) & =\frac{1}{2}\left\|\bar{E}^{\mathrm{mea}}-\bar{E}^{\mathrm{prd}}\right\|_2^2+\lambda \frac{\|\bar{n}\|_1}{\|\bar{p}\|_2} \\
& +\langle\bar{u}, \bar{\bar{D}} \bar{\tau}-\bar{n}\rangle+\frac{\rho_1}{2}\|\bar{\bar{D}} \bar{\tau}-\bar{n}\|_2^2 +\langle\bar{q}, \bar{\bar{D}} \bar{\tau}-\bar{p}\rangle+\frac{\rho_2}{2}\|\bar{\bar{D}} \bar{\tau}-\bar{p}\|_2^2
\end{aligned}
\end{equation}
where $\bar{u}$ and $\bar{q}$ are the Lagrange multipliers, and $\rho_1$ and $\rho_2$ are the penalty weights. Solving the saddle-point problem of~\eqref{augmented_lagrangian}, i.e., minimizing over $\bar{\tau}$, $\bar{n}$, and $\bar{p}$ and maximizing over the multipliers $\bar{u}$ and $\bar{q}$, is equivalent to solving the constrained problem~\eqref{constrained_ratio_problem}, and hence the original problem~\eqref{ratio_norm_objective}. 

Based on ADMM, the optimization of~\eqref{augmented_lagrangian} is decomposed into five sub-steps, each updating one of the five variables while the others are held fixed~\cite{A scale-invariant approach for sparse signal recovery}
\begin{equation}\label{admm_substeps}
\begin{aligned}
& \bar{\tau}^{(k)}= \arg\min_{\bar{\tau}} L ( \bar{\tau}, \bar{n}^{(k-1)}, \bar{p}^{(k-1)}, \bar{u}^{(k-1)}, \bar{q}^{(k-1)} ) \\
& \bar{p}^{(k)}= \arg\min_{\bar{p}} L ( \bar{\tau}^{(k)}, \bar{n}^{(k-1)}, \bar{p}, \bar{u}^{(k-1)}, \bar{q}^{(k-1)} ) \\
& \bar{n}^{(k)}= \arg\min_{\bar{n}} L ( \bar{\tau}^{(k)}, \bar{n}, \bar{p}^{(k)}, \bar{u}^{(k-1)}, \bar{q}^{(k-1)} ) \\
& \bar{q}^{(k)}= \bar{q}^{(k-1)}+ \rho_{2}( \bar{\bar{D}}\bar{\tau}^{(k)} -\bar{p}^{(k)} ) \\
& \bar{u}^{(k)}= \bar{u}^{(k-1)}+ \rho_{1}( \bar{\bar{D}}\bar{\tau}^{(k)} -\bar{n}^{(k)} ).   
\end{aligned}
\end{equation}
The first sub-step of~\eqref{admm_substeps} yields a nonlinear optimization problem for updating $\bar{\tau}$ because $\bar{E}^{\mathrm{prd}}$ is a nonlinear function of $\bar{\tau}$. Therefore, the Gauss--Newton method is used to solve this problem. The update of $\bar{\tau}$ is obtained by solving
\begin{equation}\label{gauss_newton_update}
\begin{aligned}
& \left[\Re\{\bar{\bar{J}}^{(k-1)\dagger} \bar{\bar{J}}^{(k-1)}\}+(\rho_1+\rho_2) \bar{\bar{D}}^{\top} \bar{\bar{D}}\right] \Delta \bar{\tau}\\
&= {\left[\Re \{\bar{\bar{J}}^{(k-1)\dagger} \Delta \bar{d}\}-\bar{\bar{D}}^{\top}(\bar{u}^{(k-1)}+\bar{q}^{(k-1)})\right.} \\
& \left.-(\rho_1+\rho_2) \bar{\bar{D}}^{\top} \bar{\bar{D}}\bar{\tau}^{(k-1)}+\bar{\bar{D}}^{\top}(\rho_1 \bar{n}^{(k-1)}+\rho_2 \bar{p}^{(k-1)})\right]
\end{aligned}
\end{equation}
so that $\bar{\tau}^{(k)} = \bar{\tau}^{(k-1)} + \Delta \bar{\tau}$. In~\eqref{gauss_newton_update}, $\bar{\bar{J}}^{(k-1)}$ is the Jacobian matrix, ``$\dagger$'' represents the conjugate transpose, $\Re \{\cdot\}$ is the real part operator, and $\Delta \bar{d}$ is the difference between the measured data and the predicted data at iteration $k-1$. The Jacobian is assembled as $\bar{\bar{J}} = [ \bar{\bar{J}}_{1}^{\top}, \bar{\bar{J}}_{2}^{\top}, \ldots , \bar{\bar{J}}_{N_{\mathrm{t}}}^{\top} ]^{\top}$, where $\bar{\bar{J}}_i$ corresponds to transmitter $i$ and is evaluated at $\bar{\tau}^{(k-1)}$. The Jacobian and its conjugate transpose are given by~\cite{Microwave imaging-complex permittivity reconstruction with a Levenberg-Marquardt method}
\begin{equation}\label{jacobian_matrix}
\begin{aligned}
\bar{\bar{J}}_{i} &= \bar{\bar{H}}(\bar{\bar{I}}+\mathrm{diag}\{\bar{\tau}\} \bar{\bar{G}})^{-1} \mathrm{diag}\{\bar{E}^{\mathrm{tot}}_{i}\} \\
\bar{\bar{J}}_{i}^{\dagger} &= \mathrm{diag}\{\bar{E}^{\mathrm{tot}*}_{i}\}(\bar{\bar{I}}+\bar{\bar{G}}^{\dagger}\mathrm{diag}\{\bar{\tau}^{*}\})^{-1} \bar{\bar{H}}^{\dagger}
\end{aligned}
\end{equation}
where ``$*$'' denotes the complex conjugate. In this work, the Jacobian matrix is formed explicitly so that~\eqref{gauss_newton_update} can be solved efficiently by Cholesky factorization. To reduce the computational cost, $\bar{\bar{J}}_i^{\dagger}$ is computed first: applying $(\bar{\bar{I}}+\bar{\bar{G}}^{\dagger}\mathrm{diag}\{\bar{\tau}^{*}\})^{-1}$ to the $N_{\mathrm{r}}$ columns of $\bar{\bar{H}}^{\dagger}$ requires $N_{\mathrm{r}}$ linear solves, whereas assembling $\bar{\bar{J}}_i$ directly would require one solve per cell. Since $N_{\mathrm{r}} \ll N$, the former is substantially cheaper, and $\bar{\bar{J}}_i$ is then obtained from $\bar{\bar{J}}_i^{\dagger}$ at negligible cost.

The second sub-step for $\bar{p}$ is first simplified to
\begin{equation}\label{p_subproblem}
\bar{p}^{(k)}=\arg \min _{\bar{p}} \frac{c^{(k-1)}}{\|\bar{p}\|_2}+\frac{\rho_2}{2}\left\|\bar{e}^{(k-1)}-\bar{p}\right\|_2^2
\end{equation}
by letting $c^{(k-1)}=\lambda\left\|\bar{n}^{(k-1)}\right\|_1$ and $\bar{e}^{(k-1)}=\bar{\bar{D}} \bar{\tau}^{(k)}+\frac{\bar{q}^{(k-1)}}{\rho_2}$. Depending on the value of $\bar{e}^{(k-1)}$, \eqref{p_subproblem} is solved in two cases. When $\bar{e}^{(k-1)}=\bar{0}$, the stationarity condition of~\eqref{p_subproblem} reduces to
\begin{equation}\label{zero_p_condition}
\frac{c^{(k-1)}}{\|\bar{p}^{(k)}\|_2^3}=\rho_2.
\end{equation}
So any vector satisfying $\|\bar{p}^{(k)}\|_2^2=(c^{(k-1)} / \rho_2)^{\frac{2}{3}}$ is a solution of~\eqref{p_subproblem}. This degenerate case does not arise in practice. In the general case $\bar{e}^{(k-1)} \neq \bar{0}$,  the stationarity condition becomes
\begin{equation}\label{p_cubic_relation}
(-\frac{c^{(k-1)}}{\|\bar{p}^{(k)}\|_2^3}+\rho_2) \bar{p}^{(k)}=\rho_2 \bar{e}^{(k-1)}.
\end{equation}
The solution of~\eqref{p_cubic_relation} can be shown to take the form 
\begin{equation}\label{p_scaling}
\bar{p}^{(k)}=\alpha \bar{e}^{(k-1)}
\end{equation}
\begin{equation}\label{alpha_relation}
\alpha=\frac{\rho_2} {(-\frac{c^{(k-1)}}{\|\bar{p}^{(k)}\|_2^3}+\rho_2)}.
\end{equation}
Since the minimizer $\bar{p}^{(k)}$ is aligned with $\bar{e}^{(k-1)}$, $\alpha>0$ and $\|\bar{p}^{(k)}\|_2^3=\alpha^3\|\bar{e}^{(k-1)}\|_2^3$. Substituting~\eqref{p_scaling} into~\eqref{alpha_relation} then converts the update of $\bar{p}$ into a cubic equation in $\alpha$
\begin{equation}\label{alpha_cubic}
\alpha^3-\alpha^2-\frac{c^{(k-1)}}{\rho_2\left\|\bar{e}^{(k-1)}\right\|_2^3}=0.
\end{equation}
For $\gamma>0$, \eqref{alpha_cubic} has a single real root, given by
\begin{equation}\label{alpha_root}
\alpha=\frac{1}{3}+\frac{1}{3}(\beta+\frac{1}{\beta})
\end{equation}
where
\begin{equation*}
\beta =  \sqrt[3]{\left(27 \gamma  +  2  + \sqrt{(27 \gamma + 2)^2 - 4}\right) / 2}
\end{equation*}
and $\gamma=c^{(k-1)} /(\rho_2\|\bar{e}^{(k-1)}\|_2^3)$. Thus, after solving~\eqref{alpha_cubic}, $\bar{p}$ is determined by~\eqref{p_scaling}.

The optimization of the third sub-step is given by
\begin{equation}\label{n_subproblem}
\!\!\bar{n}^{(k)}=\arg \min _{\bar{n}} \frac{\lambda \|\bar{n}\|_1}{\left\|\bar{p}^{(k)}\right\|_2}+\frac{\rho_1}{2}\left\|\bar{\bar{D}} \bar{\tau}^{(k)}+\frac{\bar{u}^{(k-1)}}{\rho_1}-\bar{n}\right\|_2^2.
\end{equation}
Equation~\eqref{n_subproblem} cannot be optimized by directly evaluating its gradient because $\ell_1$-norm is not differentiable at the origin. However, it can be efficiently handled by the soft-thresholding scheme, so that the solution of~\eqref{n_subproblem} is given by
\begin{equation}\label{shrinkage_update}
\bar{n}^{(k)}=\mathrm{shrink}\left(\bar{\bar{D}} \bar{\tau}^{(k)}+\frac{\bar{u}^{(k-1)}}{\rho_1}, \frac{\lambda}{\rho_1\left\|\bar{p}^{(k)}\right\|_2}\right)
\end{equation}
where $\mathrm{shrink}(\bar{x}, z) = \mathrm{sign}(\bar{x}) \odot \max(|\bar{x}| - z, 0)$ is the soft-thresholding operator and ``$\odot$'' denotes elementwise multiplication~\cite{Distributed optimization and statistical learning via the alternating direction method of multipliers}.

Finally, in the last two sub-steps, $\bar{u}$ and $\bar{q}$ are updated using gradient ascent, with $\rho_1$ and $\rho_2$ as the respective step sizes~\cite{Distributed optimization and statistical learning via the alternating direction method of multipliers}.

\subsection{Multi-frequency Inversion}\label{multifrequency_inversion}

In this paper, the frequency-hopping scheme is used for multi-frequency inversion~\cite{A frequency-hopping approach for microwave imaging of large inhomogeneous bodies}. At the first frequency, the initial guess of the inversion is set to $\bar{\tau}^{(0)}=\bar{0}$. The inversion result at each frequency is used as the initial guess at the next frequency. At each frequency, the solution is obtained by iteratively applying~\eqref{admm_substeps}. Algorithm~\ref{ratio_norm_inversion} summarizes the workflow of the proposed inversion scheme with $\ell_1 / \ell_2$-norm regularization. Here, $N_\mathrm{f}$ is the number of frequencies and $N_\mathrm{iter}$ is the maximum number of iterations at each frequency.

\begin{figure}[!t]
\hrule\vspace{2pt}
\refstepcounter{manuscriptalgorithm}\label{ratio_norm_inversion}
\noindent\textbf{Algorithm \themanuscriptalgorithm} $\ell_1 / \ell_2$-norm-regularized Inversion\par
\vspace{2pt}\hrule\vspace{2pt}
\noindent \, 1: \textbf{for} $\begin{aligned}f & = 1,2,\ldots,N_{\mathrm{f}}\end{aligned}$\par
\vspace{1pt}
\noindent \, 2: \quad \textbf{if} $f = 1$\par
\vspace{1pt}
\noindent \, 3: \quad\quad $\bar{\tau}^{(0)}=\bar{0}$\par
\vspace{1pt}
\noindent \, 4: \quad \textbf{else} \par
\vspace{1pt}
\noindent \, 5: \quad \quad  $\bar{\tau}^{(0)}$ is the result at frequency $f-1$\par
\vspace{1pt}
\noindent \, 6: \quad \textbf{end if}\par
\vspace{1pt}
\noindent \, 7: \quad initialize $\bar{n}^{(0)} = \bar{\bar{D}} \bar{\tau}^{(0)}$, $\bar{p}^{(0)} = \bar{\bar{D}} \bar{\tau}^{(0)}$\par
\vspace{1pt}
\noindent \, 8: \quad initialize $\bar{u}^{(0)}=\bar{0}$, $\bar{q}^{(0)}=\bar{0}$\par
\vspace{1pt}
\noindent \, 9: \quad calculate $\bar{\bar{G}}$, $\bar{\bar{H}}$, $\bar{E}^{\mathrm{inc}}_{i}$ ($i=1,2,\ldots, N_{\mathrm{t}}$)\par
\vspace{1pt}
\noindent 10: \quad calculate $\bar{E}^{\mathrm{tot}, (0)}_{i}$, $\bar{E}^{\mathrm{prd}, (0)}$ by solving the forward \par 
\noindent \quad \quad \; problem\par
\vspace{1pt}
\noindent 11: \quad \textbf{for} $k = 1,2,\ldots,N_{\mathrm{iter}}$\par
\vspace{1pt}
\noindent 12: \quad\quad calculate $\bar{\tau}^{(k)}$ using~\eqref{gauss_newton_update}\par
\vspace{1pt}
\noindent 13: \quad\quad calculate $\bar{E}^{\mathrm{tot}, (k)}_{i}$, $\bar{E}^{\mathrm{prd}, (k)}$ by solving the forward \par 
\noindent \quad \quad \quad \; problem\par
\vspace{1pt}
\noindent 14: \quad\quad \textbf{if} $\bar{e}^{(k-1)}=\bar{0}$\par
\vspace{1pt}
\noindent 15: \quad\quad\quad $\bar{p}^{(k)}$ satisfies $\|\bar{p}^{(k)}\|_2^2=(c^{(k-1)} / \rho_2)^{\frac{2}{3}}$\par
\vspace{1pt}
\noindent 16: \quad\quad \textbf{else}\par
\vspace{1pt}
\noindent 17: \quad\quad\quad calculate $\alpha$ using~\eqref{alpha_root}, then calculate $\bar{p}^{(k)}$ \par 
\noindent \quad\quad\quad \quad \; using~\eqref{p_scaling}\par
\vspace{1pt}
\noindent 18: \quad\quad \textbf{end}\par
\vspace{1pt}
\noindent 19: \quad\quad calculate $\bar{n}^{(k)}$ using~\eqref{shrinkage_update}\par
\vspace{1pt}
\noindent 20: \quad\quad $\bar{q}^{(k)}= \bar{q}^{(k-1)}+ \rho_{2}( \bar{\bar{D}}\bar{\tau}^{(k)} -\bar{p}^{(k)} )$\par
\vspace{1pt}
\noindent 21: \quad\quad $\bar{u}^{(k)}= \bar{u}^{(k-1)}+ \rho_{1}( \bar{\bar{D}}\bar{\tau}^{(k)} -\bar{n}^{(k)} )$\par
\vspace{1pt}
\noindent 22: \quad \textbf{end for}\par
\vspace{1pt}
\noindent 23: \textbf{end for}\par
\vspace{1pt}
\vspace{2pt}\hrule
\end{figure}

\section{Numerical examples}\label{numerical_examples}

In this section, the proposed regularization method is applied to the inversion of synthetic and experimental data. It is compared with two well-established methods, Tikhonov and TV regularization, and its convergence, robustness, and efficiency are analyzed. The Tikhonov and TV inversions are also built on the Gauss--Newton framework: Tikhonov is implemented following~\cite{Overview and classification of some regularization techniques for the Gauss-Newton inversion method applied to inverse scattering problems}, and TV is constructed using the same ADMM and soft-thresholding scheme as the proposed method~\cite{Distributed optimization and statistical learning via the alternating direction method of multipliers}. In all the numerical examples, the penalty weight of the TV inversion, $\rho_{\mathrm{TV}}$, is fixed as $\rho_{\mathrm{TV}}= \rho_1 + \rho_2$, so that its total gradient-penalty weight matches that of the proposed method for a fair comparison. 

To quantitatively assess the different methods, the normalized relative errors of the model, $\mathrm{NRE}_{\varepsilon}$, and the data, $\mathrm{NRE}_{\mathrm{E}}$, are used. They quantify how well the model is reconstructed and the data is fitted, and are defined as
\begin{equation}\label{model_nre}
\mathrm{NRE}_{\varepsilon}=\frac{\|\bar{\varepsilon}^{\mathrm{inv}}_{\mathrm{r}}-\bar{\varepsilon}^{\mathrm{true}}_{\mathrm{r}}\|_2}{\|\bar{\varepsilon}^{\mathrm{true}}_{\mathrm{r}}\|_2}
\end{equation}
\begin{equation}\label{data_nre}
\mathrm{NRE}_{\mathrm{E}}=\frac{\|\bar{E}^{\mathrm{prd}}-\bar{E}^{\mathrm{mea}}\|_2}{\|\bar{E}^{\mathrm{mea}}\|_2}
\end{equation}
where $\bar{\varepsilon}^{\mathrm{inv}}_{\mathrm{r}}$ and $\bar{\varepsilon}^{\mathrm{true}}_{\mathrm{r}}$ are the reconstructed and true relative permittivity.

\subsection{Synthetic Data Inversion}\label{synthetic_inversion}

For simplicity, all the models used for synthetic data inversion share the same system configuration, with $N_\mathrm{t}=16$ transmitters and $N_\mathrm{r}=16$ receivers operating at $f = \{100,200,300\}\,\mathrm{MHz}$ around the $10\,\mathrm{m} \times 10\,\mathrm{m}$ investigation domain $S$. $S$ spans $3.33 \lambda_{0} \times 3.33 \lambda_{0}$ at $100 \, \mathrm{MHz}$ and $10 \lambda_{0} \times 10 \lambda_{0}$ at $300 \, \mathrm{MHz}$. $S$ is meshed into $100 \times 100$ cells for both the forward and inverse problems. The frequency-hopping scheme with $N_{\mathrm{iter}}=10$ iterations at each frequency is used. 

\begin{figure}[!ht]
\centering
\includegraphics[width=0.5\columnwidth]{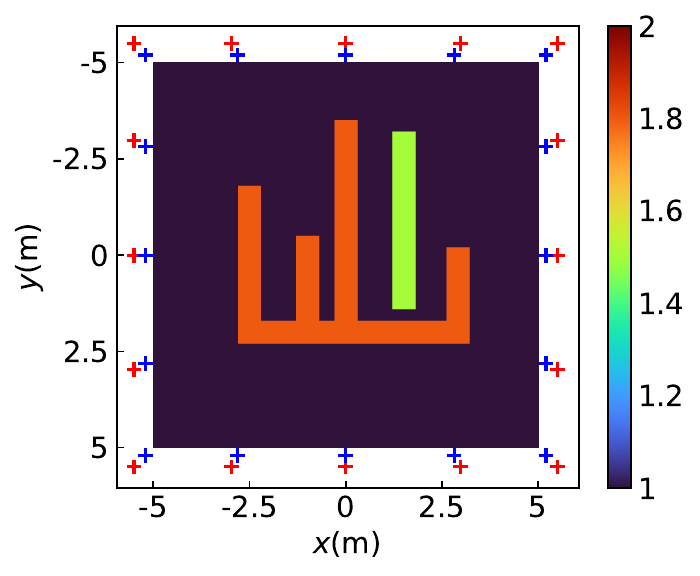}
\caption{True relative permittivity and system configuration of the first synthetic model (red +: transmitters; blue +: receivers).}\label{synthetic_model_i}
\end{figure}

\begin{figure}[!ht]
\centering
\includegraphics[width=\columnwidth]{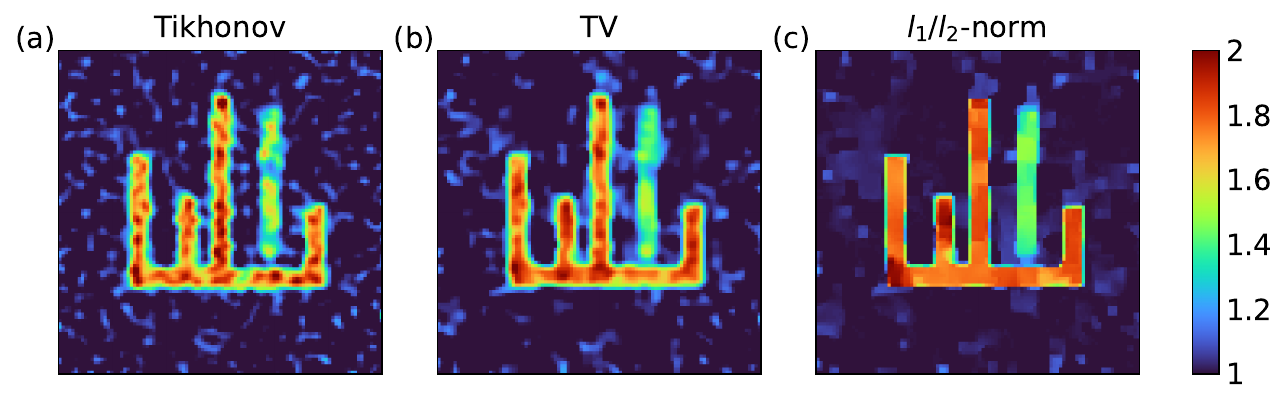}
\caption{Relative permittivity reconstructed by (a) Tikhonov, (b) TV, (c) $\ell_1 / \ell_2$-norm regularization.}\label{synthetic_model_i_results}
\end{figure}

\begin{figure}[!ht]
\centering
\includegraphics[width=0.8\columnwidth]{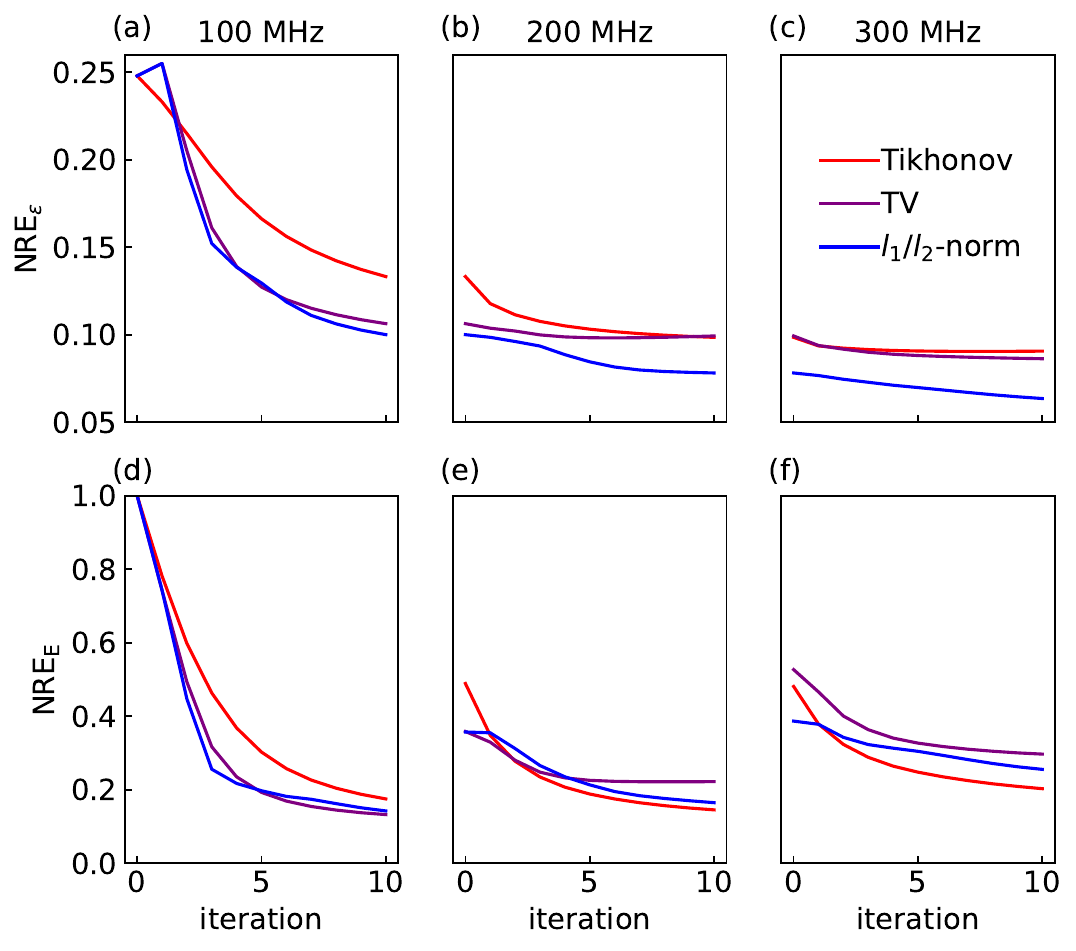}
\caption{Convergence curves of $\mathrm{NRE}_{\varepsilon}$ at (a) $100 \, \mathrm{MHz}$, (b) $200 \, \mathrm{MHz}$, (c) $300 \, \mathrm{MHz}$; and $\mathrm{NRE}_{\mathrm{E}}$ at (d) $100 \, \mathrm{MHz}$, (e) $200 \, \mathrm{MHz}$, (f) $300 \, \mathrm{MHz}$.}\label{synthetic_model_i_convergence}
\end{figure}

The relative permittivity of the first synthetic model, $\bar{\varepsilon}^{\mathrm{true}}_{\mathrm{r}}$, is shown in Fig.~\ref{synthetic_model_i}. The measured data is obtained by adding Gaussian noise to the scattered fields computed by solving the forward problem with $\bar{\varepsilon}^{\mathrm{true}}_{\mathrm{r}}$, at a signal-to-noise ratio (SNR) of $20\,\mathrm{dB}$. Fig.~\ref{synthetic_model_i_results} shows the reconstructed relative permittivity, $\bar{\varepsilon}^{\mathrm{inv}}_{\mathrm{r}}$, obtained by Tikhonov, TV, and $\ell_1 / \ell_2$-norm regularization. Tikhonov provides only a blurred reconstruction degraded by artifacts because of its over-smoothing effect [Fig.~\ref{synthetic_model_i_results}(a)]. TV improves the reconstruction by preserving the edges and recovering the piecewise-constant permittivity of the scatterers [Fig.~\ref{synthetic_model_i_results}(b)]. The $\ell_1 / \ell_2$-norm regularization performs best [Fig.~\ref{synthetic_model_i_results}(c)]: the artifacts are greatly reduced and the scatterers are well reconstructed. The quantitative improvement is confirmed by the convergence curves in Fig.~\ref{synthetic_model_i_convergence}, which show $\mathrm{NRE}_{\varepsilon}$ and $\mathrm{NRE}_{\mathrm{E}}$ at all frequencies for the different methods. The $\ell_1 / \ell_2$-norm regularization achieves the lowest $\mathrm{NRE}_{\varepsilon}$, while its $\mathrm{NRE}_{\mathrm{E}}$ remains comparable to that of the other methods, indicating that the improved reconstruction is obtained without sacrificing the data fit.

\begin{figure}[!ht]
\centering
\includegraphics[width=0.5\columnwidth]{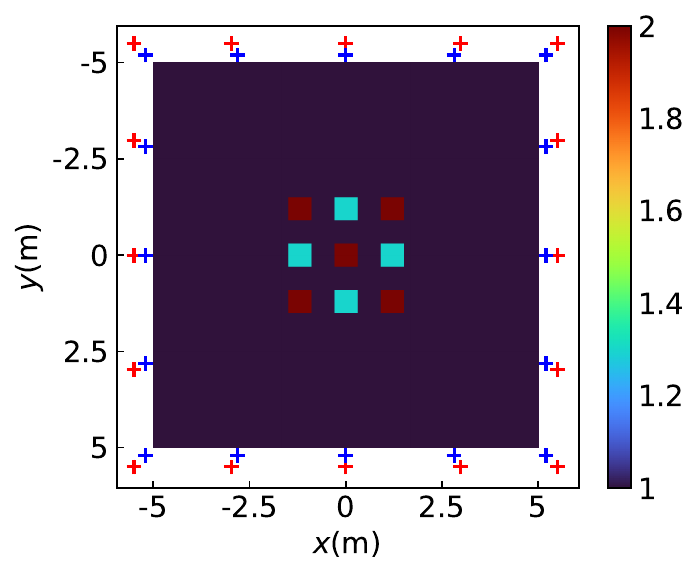}
\caption{True relative permittivity and system configuration of the second synthetic model (red +: transmitters; blue +: receivers).}\label{synthetic_model_ii}
\end{figure}

\begin{figure}[!ht]
\centering
\includegraphics[width=1\columnwidth]{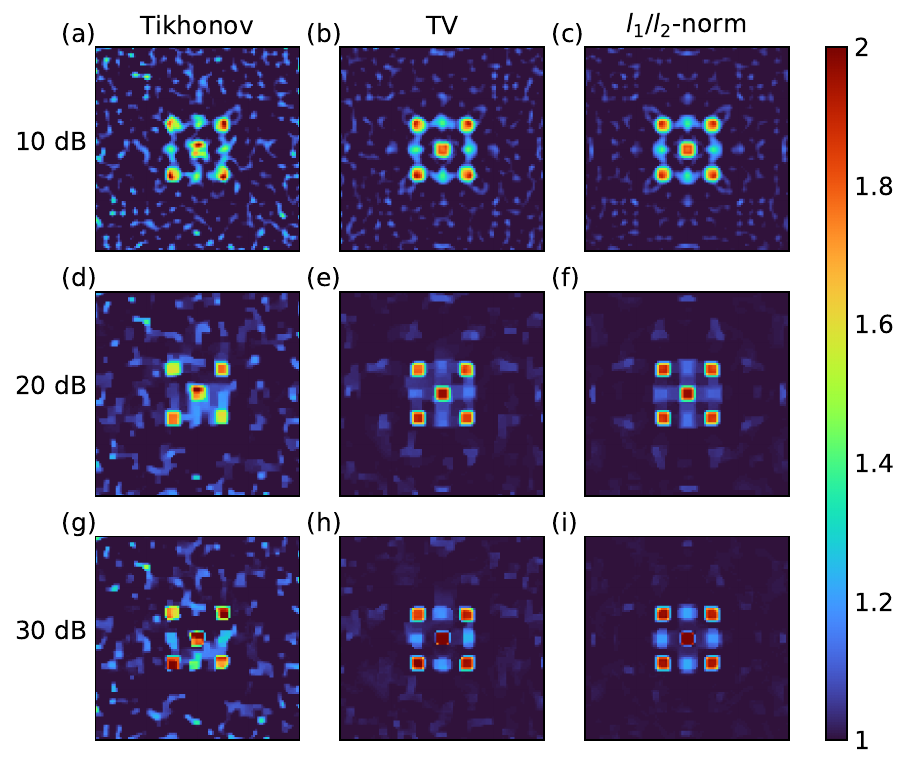}
\caption{Relative permittivity reconstructed by Tikhonov at SNR of (a) $10 \, \mathrm{dB}$, (d) $20 \, \mathrm{dB}$, (g) $30 \, \mathrm{dB}$; TV at SNR of (b) $10 \, \mathrm{dB}$, (e) $20 \, \mathrm{dB}$, (h) $30 \, \mathrm{dB}$; and $\ell_1 / \ell_2$-norm at SNR of (c) $10 \, \mathrm{dB}$, (f) $20 \, \mathrm{dB}$, (i) $30 \, \mathrm{dB}$.}\label{noise_results}
\end{figure}

\begin{figure}[!ht]
\centering
\includegraphics[width=1\columnwidth]{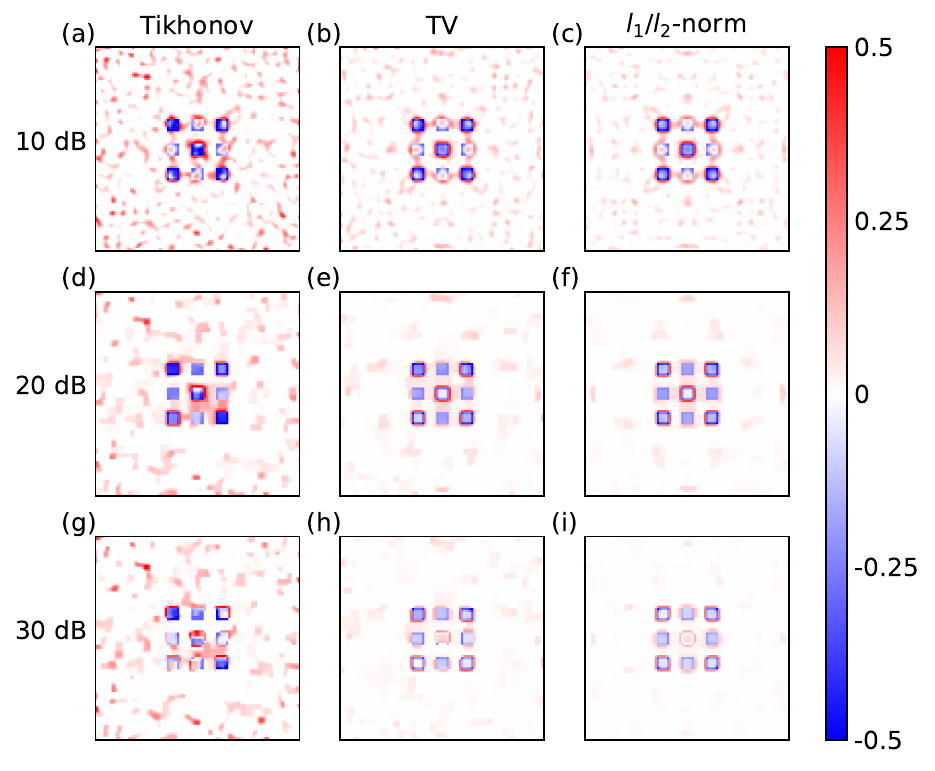}
\caption{Difference between the reconstructed and true relative permittivity, $\bar{\varepsilon}^{\mathrm{inv}}_{\mathrm{r}} - \bar{\varepsilon}^{\mathrm{true}}_{\mathrm{r}}$, obtained by Tikhonov at SNR of (a) $10 \, \mathrm{dB}$, (d) $20 \, \mathrm{dB}$, (g) $30 \, \mathrm{dB}$; TV at SNR of (b) $10 \, \mathrm{dB}$, (e) $20 \, \mathrm{dB}$, (h) $30 \, \mathrm{dB}$; and $\ell_1 / \ell_2$-norm at SNR of (c) $10 \, \mathrm{dB}$, (f) $20 \, \mathrm{dB}$, (i) $30 \, \mathrm{dB}$.}\label{noise_residuals}
\end{figure}

The relative permittivity of the second synthetic model, $\bar{\varepsilon}^{\mathrm{true}}_{\mathrm{r}}$, is shown in Fig.~\ref{synthetic_model_ii}. The three sets of measured data are obtained by adding Gaussian noise to the scattered fields computed by solving the forward problem with $\bar{\varepsilon}^{\mathrm{true}}_{\mathrm{r}}$, at SNRs of $10$, $20$, and $30 \, \mathrm{dB}$. Fig.~\ref{noise_results} shows the reconstructed relative permittivity, $\bar{\varepsilon}^{\mathrm{inv}}_{\mathrm{r}}$, obtained by Tikhonov, TV, and $\ell_1 / \ell_2$-norm regularization, and Fig.~\ref{noise_residuals} shows the corresponding difference between the reconstructed and true relative permittivity, $\bar{\varepsilon}^{\mathrm{inv}}_{\mathrm{r}} - \bar{\varepsilon}^{\mathrm{true}}_{\mathrm{r}}$. As in the previous example, Tikhonov yields over-smoothed reconstructions in which the scatterers are distorted [Fig.~\ref{noise_results}(a), (d), (g)]. TV performs better in preserving the edges [Fig.~\ref{noise_results}(b), (e), (h)]. However, because of the scale-variant nature of $\ell_1$-norm, the permittivity values recovered by TV are underestimated and the weak scatterers are over-flattened [Fig.~\ref{noise_residuals}(b), (e), (h)]. In contrast, the $\ell_1 / \ell_2$-norm does not suffer from this issue: Fig.~\ref{noise_results}(c), (f), (i) and the corresponding differences in Fig.~\ref{noise_residuals}(c), (f), (i) show that it reconstructs the permittivity profiles more accurately at every noise level. The quantitative results are given in Table~\ref{noise_metrics}. For the noisier data, $\ell_1 / \ell_2$-norm regularization avoids over-fitting the noise, which leads to a higher $\mathrm{NRE}_{\mathrm{E}}$ than the other two methods. For the less noisy data, $\ell_1 / \ell_2$-norm regularization fits the data better than the other two methods. This demonstrates that the proposed method is more robust to noise.

Note that, because of the nonconvexity and nonlinearity of $\ell_1 / \ell_2$-norm, the optimal $\lambda$ cannot be known a priori. Instead, $\lambda$ is determined here through experimental experiments. The following example illustrates that the choice of $\lambda$ is less critical for $\ell_1 / \ell_2$-norm regularization than for TV.

\begin{figure}[!ht]
\centering
\includegraphics[width=0.5\columnwidth]{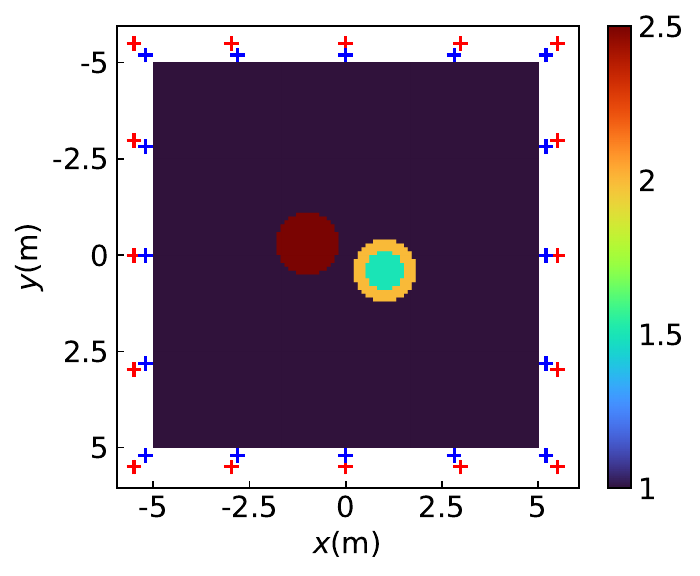}
\caption{True relative permittivity and system configuration of the third synthetic model (red +: transmitters; blue +: receivers).}\label{synthetic_model_iii}
\end{figure}

\begin{figure}[!ht]
\centering
\includegraphics[width=1.0\columnwidth]{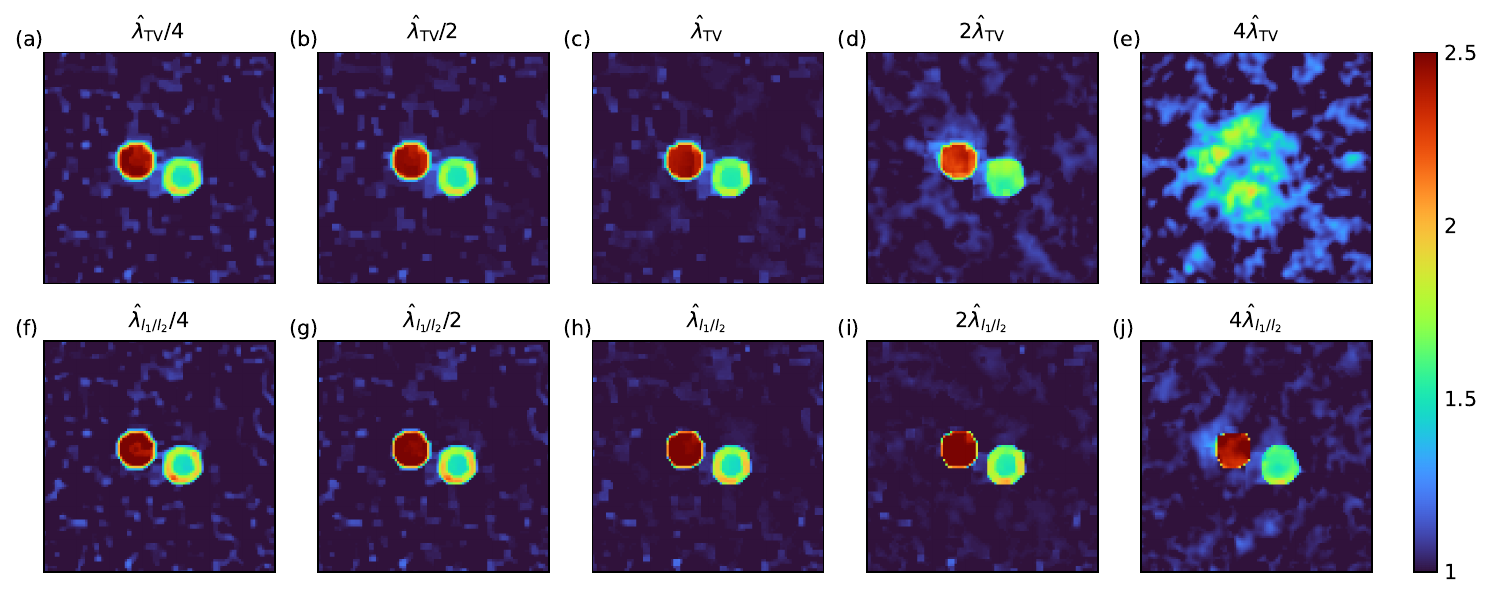}
\caption{Relative permittivity reconstructed by TV with $\lambda_{\mathrm{TV}} = $ (a) $1$, (b) $2$, (c) $4$, (d) $8$, (e) $16$; and $\ell_1 / \ell_2$-norm with $\lambda_{\ell_1 / \ell_2} = $ (f) $5$, (g) $10$, (h) $20$, (i) $40$, (j) $80$.}\label{regularization_weight_results}
\end{figure}

The relative permittivity of the third synthetic model, $\bar{\varepsilon}^{\mathrm{true}}_{\mathrm{r}}$, is shown in Fig.~\ref{synthetic_model_iii}. The measured data is obtained by adding Gaussian noise to the scattered fields computed by solving the forward problem with $\bar{\varepsilon}^{\mathrm{true}}_{\mathrm{r}}$, at an SNR of $20\,\mathrm{dB}$. The optimal weights for the two methods, $\hat{\lambda}_{\mathrm{TV}} = 4$ for TV and $\hat{\lambda}_{\ell_{1} / \ell_{2}} = 20$ for $\ell_1 / \ell_2$-norm, are determined by numerical experiments. Each method is tested over a range of $\lambda$, $\lambda_{\mathrm{TV}} = \{ \hat{\lambda}_{\mathrm{TV}} / 4, \hat{\lambda}_{\mathrm{TV}} / 2, \hat{\lambda}_{\mathrm{TV}}, 2\hat{\lambda}_{\mathrm{TV}}, 4\hat{\lambda}_{\mathrm{TV}} \}$ and $\lambda_{\ell_{1} / \ell_{2}} = \{ \hat{\lambda}_{\ell_{1} / \ell_{2}} / 4, \hat{\lambda}_{\ell_{1} / \ell_{2}} / 2, \hat{\lambda}_{\ell_{1} / \ell_{2}}, 2 \hat{\lambda}_{\ell_{1} / \ell_{2}}, 4 \hat{\lambda}_{\ell_{1} / \ell_{2}} \}$. The reconstructions obtained by TV are shown in Fig.~\ref{regularization_weight_results}(a)--(e), and those obtained by $\ell_1 / \ell_2$-norm are shown in Fig.~\ref{regularization_weight_results}(f)--(j). For $\lambda_{\mathrm{TV}} \leq \hat{\lambda}_{\mathrm{TV}}$ and $\lambda_{\ell_1/\ell_2} \leq \hat{\lambda}_{\ell_1/\ell_2}$, the reconstructions obtained by the proposed method are more accurate and show sharper edges than those obtained by TV. For $\lambda_{\mathrm{TV}} > \hat{\lambda}_{\mathrm{TV}}$, the TV reconstruction degrades rapidly, and at $\lambda_{\mathrm{TV}} = 4 \hat{\lambda}_{\mathrm{TV}}$ it breaks down entirely. In contrast, for $\lambda_{\ell_1/\ell_2} > \hat{\lambda}_{\ell_1/\ell_2}$, the $\ell_1 / \ell_2$-norm reconstruction remains stable. Overall, $\ell_1 / \ell_2$-norm outperforms TV, as confirmed by the values of $\mathrm{NRE}_{\varepsilon}$ and $\mathrm{NRE}_{\mathrm{E}}$ given in Table~\ref{regularization_weight_metrics}. This shows that $\ell_1 / \ell_2$-norm achieves better reconstructions than TV without requiring a precisely tuned regularization weight.

\subsection{Experimental Data Inversion}\label{experimental_inversion}

\begin{figure}[!ht]
\centering
\includegraphics[width=0.8\columnwidth]{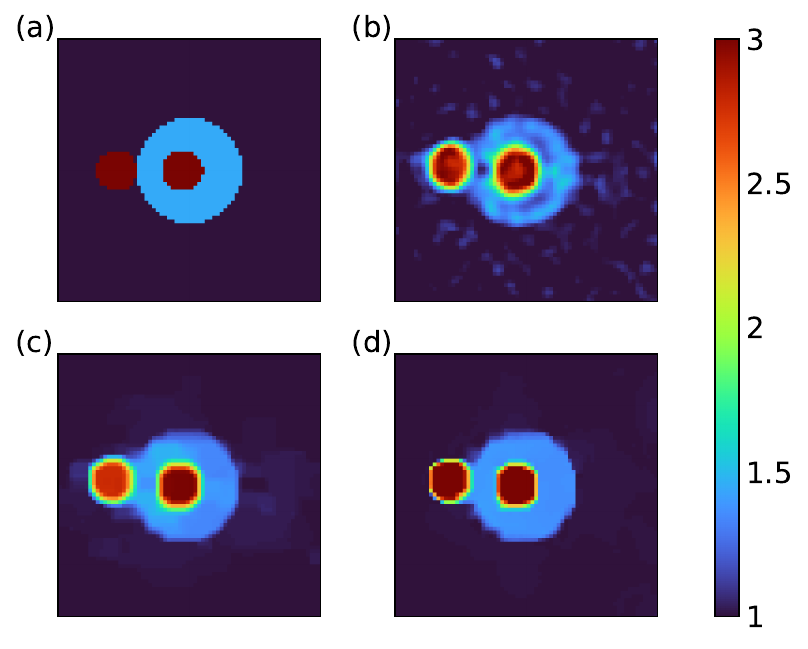}
\caption{(a) True relative permittivity of the ``FoamTwinDielTM'' model. Relative permittivity reconstructed by (b) Tikhonov, (c) TV, (d) $\ell_1 / \ell_2$-norm regularization.}\label{fresnel_results}
\end{figure}

The proposed method is further applied to the experimental dataset from Institut Fresnel to retrieve the relative permittivity of the ``FoamTwinDielTM'' model~\cite{Free space experimental scattering database continuation: Experimental set-up and measurement precision}. The model consists of one foam cylinder with radius $0.04\,\mathrm{m}$  and relative permittivity $1.45$ and two plastic cylinders with radius $0.0155\,\mathrm{m}$ and relative permittivity $3$. The relative permittivity of the model, $\bar{\varepsilon}^{\mathrm{true}}_{\mathrm{r}}$, is shown in Fig.~\ref{fresnel_results}(a). The cylinders are long enough along the $z$-axis that the inversion can be treated as a 2D TM problem. The setup consists of two horn antennas, acting as the transmitter and receiver, that rotate on a circle of radius $1.67 \, \mathrm{m}$ in the $xy$-plane. The transmitter illuminates the model at $18$ equally spaced locations, and for each location the receiver collects data at $241$ locations. The data collection is carried out at $9$ frequencies from $2 \, \mathrm{GHz}$ to $10 \, \mathrm{GHz}$. To emulate a more practical measurement, the receivers are down-sampled to $13$ per illumination, and data at $f = \{ 4, 6, 8, 10 \} \, \mathrm{GHz}$ is used in the frequency-hopping framework. The $0.2 \, \mathrm{m} \times 0.2 \, \mathrm{m}$ investigation domain is meshed into $70 \times 70$ cells. $S$ spans approximately $2.67 \lambda_{0} \times 2.67 \lambda_{0}$ at $4\,\mathrm{GHz}$ and $6.67 \lambda_{0} \times 6.67 \lambda_{0}$ at $10\,\mathrm{GHz}$.

\begin{figure}[!ht]
\centering
\includegraphics[width=0.9\columnwidth]{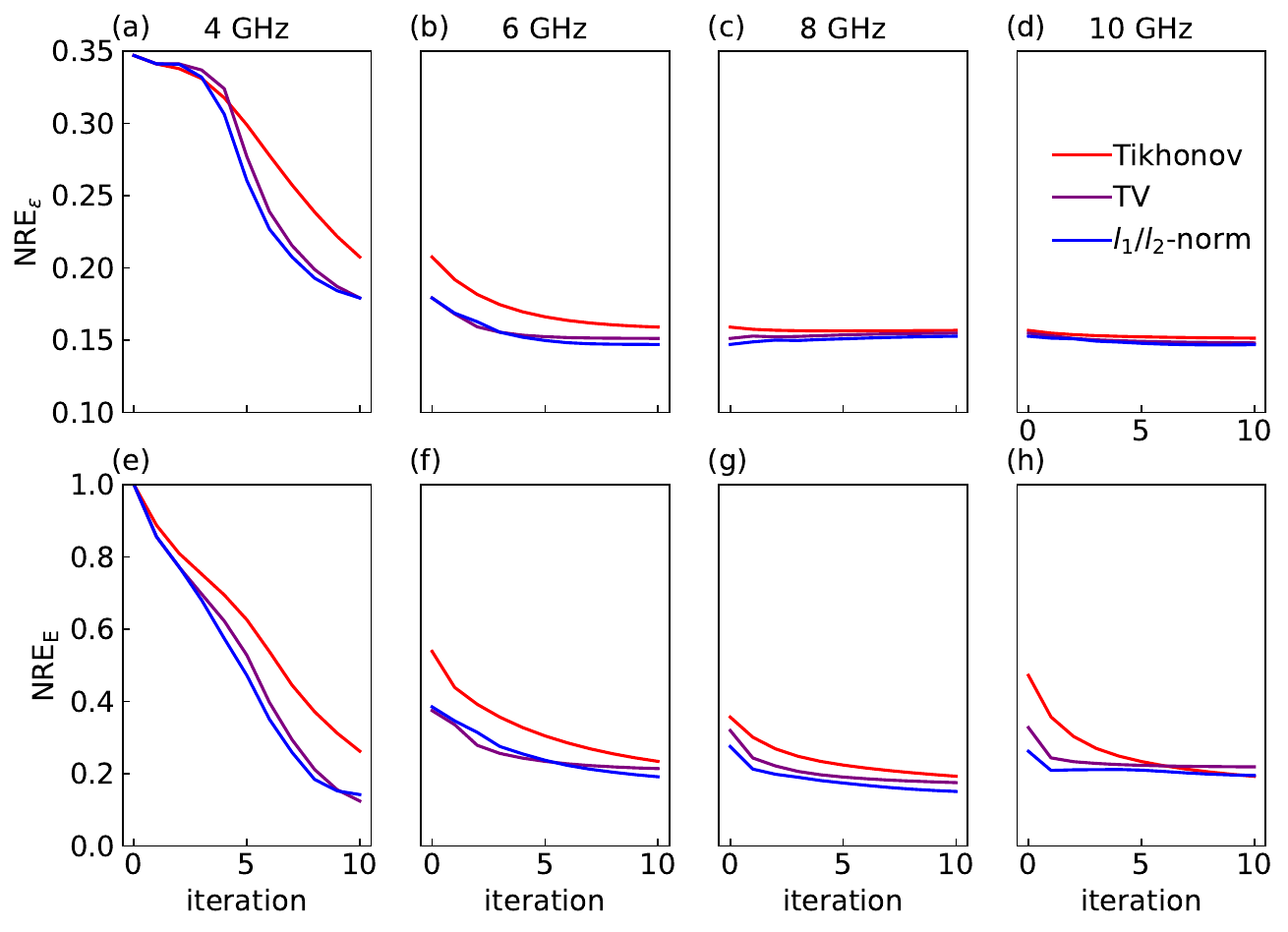}
\caption{Convergence curves of $\mathrm{NRE}_{\varepsilon}$ at (a) $4 \, \mathrm{GHz}$, (b) $6 \, \mathrm{GHz}$, (c) $8 \, \mathrm{GHz}$, (d) $10 \, \mathrm{GHz}$; and $\mathrm{NRE}_{\mathrm{E}}$ at (e) $4 \, \mathrm{GHz}$, (f) $6 \, \mathrm{GHz}$, (g) $8 \, \mathrm{GHz}$, (h) $10 \, \mathrm{GHz}$.}\label{fresnel_convergence}
\end{figure}

The reconstructions obtained by Tikhonov, TV, and $\ell_1 / \ell_2$-norm are shown in Fig.~\ref{fresnel_results}(b)--(d), respectively. The reconstruction obtained by Tikhonov is blurred, while those obtained by TV and $\ell_1 / \ell_2$-norm resolve the scatterers more clearly. The TV reconstruction has poorly defined edges and contains residual artifacts, while the $\ell_1 / \ell_2$-norm reconstruction shows sharper edges and recovers the plastic cylinders more accurately. The convergence curves in Fig.~\ref{fresnel_convergence} show that all the methods converge, with $\ell_1 / \ell_2$-norm achieving an $\mathrm{NRE}_{\varepsilon}$ comparable to or lower than those of the other methods.

Finally, the computation time of the proposed method is analyzed. As shown in Section~\ref{formulation}, the Gauss--Newton update in the first sub-step of~\eqref{admm_substeps} consumes the most time. The remaining sub-steps are computationally inexpensive. The average computation time per iteration is $7.98 \, \mathrm{s}$ for Tikhonov, $8.33 \, \mathrm{s}$ for TV, and $8.39 \, \mathrm{s}$ for the proposed method. This confirms that the $\ell_1 / \ell_2$-norm regularization adds negligible computational overhead compared with TV, since its additional sub-steps are inexpensive relative to the shared Gauss--Newton update.
 
\section{Conclusions}\label{conclusions}

In this work, a sparsity-promoting regularization based on $\ell_1 / \ell_2$-norm of the model gradient is developed to improve nonlinear EM inversion. Through a 2D visualization, the scale-invariant nature of $\ell_1/\ell_2$-norm is demonstrated. To effectively deal with the nonconvexity and nonsmoothness of $\ell_1 / \ell_2$-norm, ADMM is used to decouple its quotient structure, resulting in an equivalent problem that is solved in five sub-steps, where the Gauss--Newton method is used for the first sub-step. The proposed method is first applied to synthetic data. The reconstructions show that $\ell_1 / \ell_2$-norm is better at preserving edges and suppressing artifacts than the conventional $\ell_1$-norm. Moreover, the proposed method is robust to noise and to the choice of the regularization weight. The proposed method is then applied to experimental data, and the computation time is analyzed. The numerical examples show that $\ell_1 / \ell_2$-norm regularization provides better reconstructions than conventional methods such as Tikhonov and TV, at comparable computational cost.

\newpage\clearpage

\section*{Figures}

\newpage\clearpage

\newpage\clearpage

\newpage\clearpage

\newpage\clearpage

\newpage\clearpage

\newpage\clearpage

\newpage\clearpage

\newpage\clearpage

\newpage\clearpage

\section*{Tables}

\begin{table}[!ht]
\caption{$\mathrm{NRE}_{\varepsilon}$ and $\mathrm{NRE}_{\mathrm{E}}$ at Different Noise Levels}\label{noise_metrics}
\centering
\renewcommand{\arraystretch}{1.2}
\begin{tabular}{ccccc}
SNR (dB) & Regularization & $\mathrm{NRE}_{\varepsilon}$ & $\mathrm{NRE}_{\mathrm{E}}$ \\
\hline
10 & Tikhonov & 0.0916 & \textbf{0.5161} \\
- & TV & 0.0816 & 0.5497 \\
- & $\ell_1 / \ell_2$-norm & \textbf{0.0756} & 0.6492 \\
\hline
20 & Tikhonov & 0.0681 & 0.2328 \\
- & TV & 0.0566 & \textbf{0.2079} \\
- & $\ell_1 / \ell_2$-norm & \textbf{0.0410} & 0.2268 \\
\hline
30 & Tikhonov & 0.0646 & 0.1568 \\
- & TV & 0.0531 & 0.1424 \\
- & $\ell_1 / \ell_2$-norm & \textbf{0.0366} & \textbf{0.1290} \\
\end{tabular}
\end{table}

\begin{table}[!ht]
\caption{$\mathrm{NRE}_{\varepsilon}$ and $\mathrm{NRE}_{\mathrm{E}}$ Obtained Using Different $\lambda$}\label{regularization_weight_metrics}
\centering
\renewcommand{\arraystretch}{1.2}
\begin{tabular}{cccc}
Regularization & $\lambda$ & $\mathrm{NRE}_{\varepsilon}$ & $\mathrm{NRE}_{\mathrm{E}}$ \\
\hline
TV & 1 ($\hat{\lambda}_{\mathrm{TV}} / 4$) & 0.0693 & 0.3834 \\
- & 2 ($\hat{\lambda}_{\mathrm{TV}} / 2$) & 0.0664 & 0.3432 \\
- & 4 ($\hat{\lambda}_{\mathrm{TV}}$) & \textbf{0.0653} & \textbf{0.3188} \\
- & 8 ($2 \hat{\lambda}_{\mathrm{TV}}$) & 0.0840 & 0.6272 \\
- & 16 ($4 \hat{\lambda}_{\mathrm{TV}}$) & 0.2561 & 1.3458 \\
\hline
$\ell_1 / \ell_2$-norm & 5 ($\hat{\lambda}_{\ell_{1} / \ell_{2}} / 4$) & 0.0638 & 0.4083 \\
- & 10 ($\hat{\lambda}_{\ell_{1} / \ell_{2}} / 2$) & 0.0590 & 0.3809 \\
- & 20 ($\hat{\lambda}_{\ell_{1} / \ell_{2}}$) & \textbf{0.0561} & \textbf{0.3522} \\
- & 40 ($2 \hat{\lambda}_{\ell_{1} / \ell_{2}}$) & 0.0569 & 0.3670 \\
- & 80 ($4 \hat{\lambda}_{\ell_{1} / \ell_{2}}$) & 0.0852 & 0.6166 \\
\end{tabular}
\end{table}

\end{document}